\documentclass{article}
\usepackage{tabularx}      
\usepackage{siunitx,array}       
\usepackage{graphicx}
\usepackage{placeins}
\usepackage{subcaption}
\usepackage{amsmath}
\usepackage{booktabs}
\usepackage{geometry}
\usepackage{multirow}
\usepackage{xcolor}
\usepackage{authblk}
\usepackage{float}
\usepackage[utf8]{inputenc}      
\DeclareUnicodeCharacter{2009}{\,}   
\DeclareUnicodeCharacter{2011}{-}    
\DeclareUnicodeCharacter{2013}{--}   

\title{Transcorrelated Theory for Transition Metal Atoms}
\author[1]{Kristoffer Simula}
\author[2]{Maria-Andreea Filip}
\author[1,2]{Ali Alavi}
\affil[1]{Max Planck Institute for Solid State Research, Heisenbergstr. 1, 70569 Stuttgart, Germany}
\affil[2]{Yusuf Hamied Department of Chemistry, University of Cambridge, Lensfield Road, Cambridge CB2 1EW, United Kingdom}
\date{\today}

\begin{document}

\maketitle

\begin{abstract}  
  We benchmark ionisation and excitation energies of transition-metal atoms Sc–Zn with a transcorrelated Hamiltonian combined with pseudopotentials. The similarity transformed Hamiltonian provides compact TC wave functions in affordable aug-cc-pVTZ and aug-cc-pVQZ Gaussian bases and eliminates the need for complete basis set extrapolations. The use of Douglas–Kroll–Hess theory is omitted because scalar relativistic effects are included in the pseudopotentials. Treating the full semicore (3s 3p) valence and freezing only 1s–2p shells, we reach chemical accuracy for all atoms and properties with coupled cluster and full configuration interaction quantum Monte Carlo. Consistent total energies across disparate orbital sets and correlation solvers highlights the robustness of the TC workflow. Our study pushes benchmark-quality quantum chemistry into the 3d block without large-scale basis sets and opens a practical route for transcorrelation to strongly correlated molecules and materials hosting heavier transition metals.
\end{abstract}
\section{Introduction}

Transition-metal (TM) atoms lie at the heart of modern catalysis, magnetic materials, and emergent quantum technologies. The large configurational space in the open $3d$ shell enables multiple accessible oxidation states for redox catalysis, strong and tunable magnetism, and pronounced electronic and optical responses. The same configurational freedom makes TMs also  notoriously difficult to describe by means of theoretical quantum chemistry or physics: the delicate balance among competing spin and charge states demands methods that treat static and dynamic correlation accurately. Benchmark quality calculations on TM atoms remain indispensable for the development of new theories and methods in quantum chemistry needed to tackle the complexity of molecules and solids containing TMs.

Reaching an accurate description of TM atoms or molecules is often a central step for emerging new theories and methods in electronic structure theory. Auxiliary-field quantum Monte Carlo has in recent years succesfully modelled TMs~\cite{shee2018,shee2019,neugebauer2023}, and full configuration interaction Monte Carlo (FCIQMC) has, a few years after its formulation, shown the best ionisation potential results to date for TM atoms from scandium to zinc~\cite{thomas2015}. So far, the most accurate excitation energies for the $4s^2 3d^{m-2} \to 4s^1 3d^{m-1}$ transition of atoms from Sc to Cu have been modelled with a coupled cluster theory, taken up to the level of quadruple excitations, and with additional FCI corrections~\cite{blabanov2006}. For comparisons with experiment, all-electron (AE) methods must estimate the scalar relativistic effects. This is done usually with Douglas-Kroll-Hess (DKH) theory~\cite{peng2009} with related augmented correlation-consistent polarized basis sets~\cite{blabanov2005,blabanov2006}. Another crucial but expensive complication has been the extrapolation to the complete basis set (CBS) limit.

Transcorrelation (TC) theory is a very promising explicitly correlated approach for addressing the basis set error~\cite{luo2018,cohen2019,ammar2023,ammar2023_2,lee2023,ten2023, ammar2024}. For first and second row atoms and molecules it has demonstrated consistent convergence to the CBS limit in computationally feasible basis sets, such as cc-pVTZ or cc-pVQZ or their augmented counterparts~\cite{haupt2023,filip2025}. Besides fast convergence to the CBS, TC compactifies the wave functions~\cite{dobrautz2019}, allowing, for example, FCI-like accuracy with lower level coupled cluster methods~\cite{schraivogel2021,schraivogel2023}, or acceleration of FCI methods, such as FCIQMC~\cite{luo2018,dobrautz2019}  and selected CI \cite{ammar2023, ammar2024}, allowing convergence to  be achieved with fewer walkers or determinants, and hence resources. The xTC approximation~\cite{christlmaier2023} has removed the need for explicit evaluation of 3-body interactions  (and therefore all six-index interaction tensors) with TC without loss of accuracy, and a recent extension to pseudopotential (PP) support\cite{simula2025} has opened up possibilities to tackle larger systems and heavier atoms, while easing the problem of Jastrow optimisation for the TC method. The inclusion of scalar relativistic effects in the pseudopotentials also allows us to omit the use of DKH theory and related basis sets.

Here we present a TC study of the properties of TM atoms Sc to Zn, employed within the PP approximation. First, we outline the electronic states that need to be accurately modelled. Then we present the theory and methods in our approach. The orbital generation with gaussian basis sets is done with Hartree-Fock (HF) or density functional theory (DFT) followed by coupled cluster calculations when evaluating ionisation potentials. Estimation of spin gap and excitation energies is done with state-averaged complete active space self consistent field (SA-CASSCF) orbitals, followed by FCIQMC simulations. In between the orbital generation and correlated simulation steps, we employ an array of methods to optimize a Jastrow factor and to do the similarity transformation on the second-quantized Hamiltonian. To account for the core-valence interactions, which are found to be crucial in determination of the properties of these systems~\cite{blabanov2005,blabanov2006,thomas2015}, we include the complete semicore space into our calculations, with the $1s$, $2s$, and $2p$ orbitals frozen within the PP. 


We find that the TC workflow with pseudopotentials provides consistently chemically accurate results in AVTZ or AVQZ basis sets for TM atom excitation and ionisation without the need for CBS extrapolation or use of DK Hamiltonians and basis sets. The use of different orbitals and correlated methods provides comparable and often identical total energies, underlining the robustness of our approach to the theory of atoms. The results thus present yet another advancement in the quantum theory of atoms but, more importantly, open up a roadmap for wider application of TC methods to large and strongly correlated systems consisting of heavy atoms.

\section{Atomic states and properties}

Third-row transition metals in the periodic table of elements include Scandium (Sc), Titanium (Ti), Vanadium (V), Chromium (Cr), Manganese (Mn), Iron (Fe), Cobalt (Co), Nickel (Ni), Copper (Cu), and Zinc (Zn). Apart from Zn and Cu, they have partially filled 3d shells and their ground-state occupation of the $4s$ and $3d$ orbitals is often ambiguous. For example, the ground states of neutral Cr, Ni, and Cu have singly occupied $4s$ orbitals, while the rest of the atoms have doubly occupied $4s$ orbitals in the ground state configuration. The electrons in the $3d$ orbitals maximize their spin alignment according to Hund's rules. 

The $4s^2 3d^{m-2} \to 4s^1 3d^{m-1}$ gaps ($m$ is the number of valence electrons) of atoms Sc, Ti, and V correspond to the difference between the lowest energies of the states with spin quantum numbers $S_z=m-2$ and $S_z=m$. For these atoms the low-$S_z$ state is lower in energy. Cr has 4s$^1$3d$^5$ ground-state configuration with $S_z=6$, and there the first excited state has configuration 4s$^2$3d$^4$ with $S_z=4$. For atoms Mn--Cu, the $4s^2 3d^{m-2}$ and $4s^1 3d^{m-1}$ configurations have the same $S_z$ and the energy gap corresponds to a direct excitation. 

For ions, the $4s$ orbital is never doubly occupied. There is zero $4s$ orbital occupation for the $+1$ ions of V, Cr, Co, Ni and Cu in the ground state, and single occupation for the rest. The ionisation energies correspond to the energy difference between the ground state of the neutral atom and the ground state of the ion. 

The electronic configurations and their lowest-energy states, expressed as term symbols, for the neutral atoms and ions, and for the first excited state of the neutral atoms, are listed in Table ~\ref{table:target_occupations}. Term symbols give the spin and orbital angular momentum quantum numbers of the states. If spin-orbit coupling is omitted from the electronic Hamiltonian, as in this study, the orbital angular momentum quantum number $L$ gives the degeneracy of a state, $2L+1$, in a simulation targeting $j$-averaged states with fixed $S_z$. Hence for example the degeneracy of the Fe ground state in an xTC-FCIQMC simulation is $2L+1=2(2)+1=5$. 

It should be noted that while the $4s^2 3d^8$ is the configuration of the Ni ground state, the excitation to $4s^1 3d^9$ from experiment has a negative sign in Table~\ref{tab:ex_ip_allbasis}, caused by the $j$-averaging of the configuration energies. We present the experimental excitation and ionisation energies, where the effects caused by $j$-averaging of states effects have been removed, in Table \ref{tab:ex_ip_allbasis}. 

\begin{table}[H]
    \centering
    \scriptsize
    \renewcommand{\arraystretch}{1.2}
    \caption{Electronic configurations and corresponding states of TM atoms studied in this work.}
    \label{table:target_occupations}
    \begin{tabular}{lcccc}
    \toprule
    \textbf{Atom} & \textbf{m} & \textbf{4s$^2$3d${^m-2}$ state (neutral)} & \textbf{4s$^1$3d${^m-1}$ state (neutral)} & \textbf{Ion state (occupation)} \\
    \midrule
    Sc & 3 & $^2$D & $^4$F & $^3$D (4s$^1$3d$^1$) \\
    Ti & 4 & $^3$F & $^5$F & $^4$F (4s$^1$3d$^2$) \\
    V & 5 & $^4$F & $^6$D & $^5$D (3d$^4$) \\
    Cr & 6 & $^5$D & $^7$S & $^6$S (3d$^5$) \\
    Mn & 7 & $^6$S & $^6$D & $^7$S (4s$^1$3d$^5$) \\
    Fe & 8 & $^5$D & $^5$F & $^6$D (4s$^1$3d$^6$) \\
    Co & 9 & $^4$F & $^4$F & $^3$F (3d$^8$) \\
    Ni & 10 & $^3$F & $^3$D & $^2$D (3d$^9$) \\
    Cu & 11 & $^2$D & $^2$S & $^1$S (3d$^{10}$) \\
    Zn & 12 & $^1$S & -- & $^2$S (4s$^1$3d$^{10}$) \\
    \bottomrule
    \end{tabular}
    \end{table}

\section{Transition metal atoms with pseudopotentials}

The latest generation of pseudopotentials for transition metal atoms Sc-Zn with gaussian basis sets has been provided by Trail and Needs~\cite{trail2017} and Annaberdiyev et al.~\cite{annaberdiyev2018}. The former have eCEPPs for atoms from Sc to Fe, and Cu, while the latter have ccECPs for the whole row from Sc to Zn. Only the $1s$, $2s$, and $2p$ orbitals are included in the core of eCEPPs and ccECPs, as the $3s$ and $3p$ orbitals have been found to be important for the properties of these systems~\cite{dolg1987,ceperley1995}. Scalar relativistic effects are included in the pseudopotentials, and hence we can omit the use of DKH theory~\cite{peng2009} for Hamiltonians and associated augmented correlation-consistent polarized basis sets~\cite{blabanov2005,blabanov2006}.

In terms of the electron affinity, ionization potential, and the neutral and first ionized $ 4s\to 3d$ transitions, the PP results have been shown to have discrepancies of $3-16$ meV for Sc-Fe (eCEPP) and $3-16$ meV for Co-Zn (ccECP) to the AE values, when the AE and PP results were extrapolated to CBS limit~\cite{annaberdiyev2018}. This enables us to target chemical accuracy in our calculations. Furthermore, as the optimization of the Jastrow factor has proved to be much easier with pseudopotentials~\cite{simula2025}, TC with PPs can perform even better than the all-electron TC calculations.

\section{Theory and methods}

\begin{figure}[H]
    \includegraphics[scale=1.]{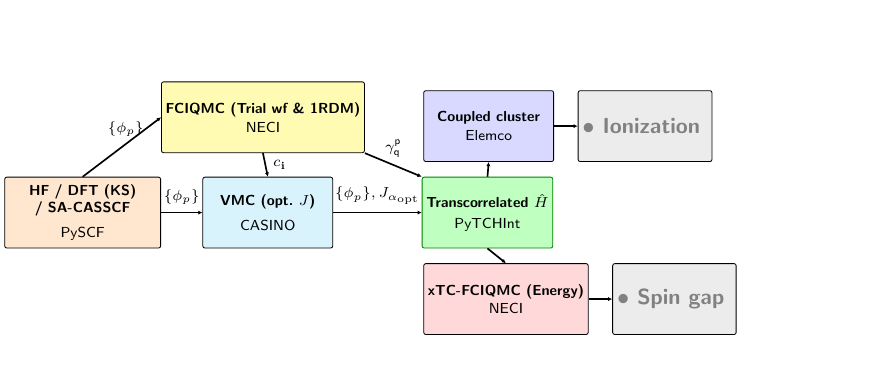}
    \caption{Computational workflow for transcorrelated theory applied to transition metals. The workflow includes orbital generation, a preliminary FCIQMC calculation, VMC optimization of a Jastrow factor, construction of the transcorrelated Hamiltonian, and advanced post-HF methods.}
    \label{fig:workflow}
\end{figure}

\subsection{Workflow}

The overall workflow involves a series of computational steps: orbital generation, a preliminary Full Configuration Interaction Quantum Monte Carlo (FCIQMC) calculation (to generate a reference wavefunction), Variational Monte Carlo (VMC) optimization of a Jastrow factor based on the aforementioned reference wavefunction, construction of the transcorrelated Hamiltonian within the xTC approximation, and finally, advanced post-Hartree-Fock (HF) methods, such as Coupled Cluster (CC) and FCIQMC. With TC Hamiltonian we denote these methods as xTC-CC and xTC-FCIQMC. The workflow is illustrated in Figure~\ref{fig:workflow}.

\subsection{Orbital generation}

In the TM metals, the choice of single-particle representation is important when targeting a state with specific spin and ionisation level. For this purpose a variety of ways of obtaining single-particle orbitals were investigated. We discuss the reference orbitals provided by different methods later in the results-section and give here a brief overview of the single-particle methods used in this work.

Single-particle orbitals $\phi_p=\sum_\alpha C^\alpha_p \theta^\alpha$, with $\theta^\alpha$ being the basis set functions, are generated with Hartree-Fock (HF), density functional theory (DFT), or state-averaged complete active space self-consistent field (SA-CASSCF) methods. In this phase, pseudopotentials are used to replace the core electrons, and only the valence electrons are treated explicitly. 

\paragraph{Hartree--Fock (HF).}
The canonical HF orbitals \(\{\phi_p\}\) follow from
\[
\hat F\,\phi_p=\varepsilon_p\phi_p,\qquad
\hat F=-\tfrac12\nabla^{2}+V(\mathbf r)+\hat J-\hat K,
\]
where \(V\) is the external (nuclear) potential and \(\hat J\) and \(\hat K\) are the Coulomb and exchange operators.

\paragraph{Kohn--Sham density--functional theory (DFT).}
With DFT we solve local the Kohn–Sham equations
\[
\Bigl[-\tfrac12\nabla^{2}+v_{\text{eff}}(\mathbf r)\Bigr]\phi_p(\mathbf r)=\varepsilon_p\phi_p(\mathbf r),
\]
with
\[
v_{\text{eff}}(\mathbf r)=V(\mathbf r)+v_{\text H}[\rho](\mathbf r)+v_{\text{xc}}[\rho](\mathbf r),\qquad
v_{\text H}[\rho]=\int\frac{\rho(\mathbf r')}{|\mathbf r-\mathbf r'|}\,d\mathbf r'.
\]
Here \(\rho(\mathbf r)=\sum_p|\phi_p(\mathbf r)|^{2}\) and
\(
v_{\text{xc}}=\delta E_{\text{xc}}[\rho]/\delta\rho
\)
derives from an explicit exchange–correlation functional \(E_{\text{xc}}[\rho]\), the defining ingredient of DFT. We use PBE \cite{perdew1996} and a hybrid B3LYP \cite{becke1993} functionals in this work. 

\paragraph{State-averaged CASSCF.}
Within a selected active space, the multiconfigurational wave function
\[
|\Psi_I^{S}\rangle=\sum_\mathbf{i} c_\mathbf{i}\,|D_\mathbf{i}\rangle
\]
is optimized by enforcing stationarity of the total energy with respect to both the molecular-orbital coefficients \(C_p^\alpha\) and the CI coefficients \(c_\mathbf{i}\):
\[
\frac{\partial E}{\partial C_p^\alpha}=0,\qquad
\frac{\partial E}{\partial c_\mathbf{i}}=0.
\]
We adopt a state-averaged scheme, evaluating energies and gradients with density matrices averaged over the targeted spin states \(\{|\Psi_I^{S}\rangle\}\). The state-averaging is done over all possible states with the same spin multiplicity and number of electrons in the $3d$ and $4s$ orbtials, or up to $100$ states in case of large number of states.

\subsection{FCIQMC}

After generating the orbitals, we perform a preliminary FCIQMC calculation with a modest walker population, $1e6$ to $3e6$ walkers. FCIQMC stochastically  evolves the wave function of shape $\Psi(\tau) = \sum c_\textbf{i}(\tau) |D_\textbf{i}\rangle$ in imaginary time $\tau$, where $|D_\textbf{i}\rangle$ are the determinants built out of $\phi_i$ and $c_\textbf{i}$ are the determinant coefficients proportional to $n_\textbf{i}$, the populations of walkers in a given determinant $D_\textbf{i}$ during a simulation. The imaginary time evolution is governed by the equation~\cite{booth2009}
\begin{equation}
  \label{equation: fciqmc population dynamics}
    n_\textbf{i}(\tau + \Delta\tau) = n_\textbf{i}(\tau)-\Delta\tau \left[\sum_{\textbf{i}\neq\textbf{j}}\hat{H}_\textbf{ij}n_\textbf{j}(\tau)\right] - \Delta\tau\left( \hat{H}_\textbf{ii} - S\right)n_\textbf{i}(\tau)
\end{equation}
where $\hat{H}_\textbf{ij}=\langle D_\textbf{i} | \hat{H} | D_\textbf{j}\rangle$ contains the matrix elements of the electronic Hamiltonian in determinant space and $S$ is the shift parameter used in population control and, in some cases, energy estimation, see later. $\tau$ is imaginary time, and $\Delta\tau$ is the time step.

An FCIQMC calculation is started from a single determinant, corresponding to a given ground- or excited state, and proceeds then with an iterative algorithm consisting of steps called ``spawning", ``death", and ``annihilation" to simulate dynamics determined by Eq.~\ref{equation: fciqmc population dynamics}~\cite{booth2009}. For proposing new spawns in the algorithm we use the precomputed heat-bath (PCHB) method as described in~\cite{weser2023}. FCIQMC dynamics is divided into equilibration and accumulation phases (separated by vertical dashed line in the FCIQMC dynamics figures of Supplementary material), where in the former $\Delta\tau$ is optimized, and the shift $S$ is kept constant, causing the population of walkers to increase exponentially. After a wanted walker population is reached, the time step is fixed and the shift is allowed to vary so that the walker population remains roughly constant. In this accumulation phase expectation values of wave function properties can be extracted by averaging over the single instances of the values of properties at different imaginary time steps.  

Using  $n_\textbf{i} $ we save the coefficients $c_\textbf{i}=n_\mathbf{i}/\sqrt{\sum_\mathbf{i}n_\mathbf{i}^2}$ of the $100$ most populated determinants at the end of an accumulation phase, which are used to construct a trial wave function for VMC.  The 1-RDMs $\gamma^\text{p}_\text{q}(\tau) =  \langle \Psi(\tau) | \hat{a}_p^\dagger \hat{a}_q | \Psi(\tau)\rangle$ are also saved, using a FCIQMC technique described in Ref.~\cite{overy2014}, for evaluation of the xTC integral contributions. This technique using multideterminant trial wave functions for Jastrow optimizations, leveraging FCIQMC, is described in detail in Ref.~\cite{haupt2025}

\subsection{Variational Monte Carlo optimization of Jastrow factor}

In the VMC step we optimize a Jastrow factor $J_{\alpha_u,\alpha_\chi,\alpha_f}=\sum_{i\neq i}u_{\alpha_u}(\mathbf{r}_i,\mathbf{r}_j) + \sum_\text{I}\sum_\text{i}\chi_{\alpha_\chi}(\mathbf{r}_i,\mathbf{R}_\text{I}) + \sum_\text{I}\sum_{\text{i}\neq\text{j}}f_{\alpha_f}(\mathbf{r}_i,\mathbf{R}_\text{I},\mathbf{r}_j)$, defined by parameters $\alpha_u,\alpha_\chi,\alpha_f$, of Drummond-Towler-Needs type \cite{drummond2004} with two-body ($u$), one-body ($\chi$), and three-body ($f$) terms. In the Jastrow terms, $\mathbf{r}_i$ and $\mathbf{r}_j$ are the positions of the electrons, and $\mathbf{R}_\text{I}$ is the position of the nuclei. 

The Jastrow factor is optimized by minimising the variance of the reference energy,
\begin{equation}
  \sigma^2_\text{ref} = \sum_\mathbf{i} \left| \langle D_\mathbf{i} | \hat{H}_{\text{TC}} | \Psi_\text{MD} \rangle - \langle D_\mathbf{i} | \Psi_\text{MD} \rangle \langle \Psi_\text{MD} | \hat{H}_{\text{TC}} | \Psi_\text{MD} \rangle \right|^2,
\end{equation}
where the sum runs over the trial wave function determinants $D_\mathbf{i}$, and $\Psi_\text{MD}=\sum_{i=1}^{100} c_\mathbf{i} D_\mathbf{i}$ is the trial wave function constructed from the FCIQMC determinant coefficients of largest amplitudes, and $\hat{H}_{\text{TC}}=e^{J_\alpha}\hat H e^{-J_\alpha}$ is the transcorrelated Hamiltonian~\cite{haupt2025}.  The procedure for a single determinant trial wave function is based on the same principles, and has been described in detail in Ref.~\cite{haupt2023}, and as in there we use enough configurations to ensure that the optimizer can resolve the reference energy to $0.1$mHa. We provide the optimized Jastrow parameters in the Supplementary material.

\subsection{Transcorrelated Hamiltonian with xTC approximation and pseudopotentials}

Using the orbitals \(\{\phi_p\}\), Jastrow factor \(J_{\alpha_u,\alpha_\chi,\alpha_f}\), and the reduced density matrix \(\gamma^p_q\), we build the second-quantized transcorrelated Hamiltonian  
\begin{equation}
\hat H_{\text{TC}}
  =e^{J_\alpha}\hat H e^{-J_\alpha}
  \xrightarrow[\text{xTC-PP}]{\text{\cite{cohen2019,christlmaier2023,simula2025}}}
  \sum_{pq}F_{pq}\,\tilde a_p^\dagger\tilde a_q
 +\tfrac12\sum_{pqrs}V_{pqrs}\,\tilde a_p^\dagger\tilde a_q^\dagger\tilde a_r\tilde a_s.
\end{equation}

\paragraph{One- and two-body matrix elements.}
\begin{align}
F_{pq} &=
      h^p_q
    +\sum_{rs}\!\bigl(U_{pr}^{qs}-U_{pr}^{sq}\bigr)\gamma^r_s
    -\sum_{rtsu}\!\bigl(L_{prt}^{qsu}-L_{prt}^{squ}-L_{prt}^{usq}\bigr)
      \gamma^{rt}_{su}, \\[4pt]
V_{pqrs} &=
      U_{pr}^{qs}+P_{pr}^{qs}
    -\sum_{tu}\!\bigl(L_{prt}^{qsu}-L_{prt}^{qus}-L_{prt}^{usq}\bigr)\gamma^t_u .
\end{align}

\paragraph{Integral definitions.}
\begin{align}
    h^p_q &= \langle\phi_p|\hat h|\phi_q\rangle, \\
    U_{pr}^{qs} &= V_{pr}^{qs}-K_{pr}^{qs}, \nonumber \\[2pt]
    V_{pr}^{qs} &= \langle\phi_p\phi_r|r_{12}^{-1}|\phi_q\phi_s\rangle, \\
    K_{pr}^{qs} &= \tfrac12\!\bigl\langle\phi_p\phi_r\bigl|
                  \mathrm P_{12}^{(1)}
                  [\nabla_1^2 J_\alpha+\nabla_1 J_\alpha+2\nabla_1 J_\alpha\!\cdot\!\nabla_1]
                  \bigr|\phi_q\phi_s\bigr\rangle, \nonumber \\[4pt]
    L_{prt}^{qsu} &= \bigl\langle\phi_p\phi_r\phi_t\bigl|
                    \mathrm P_{23}^{(1)}
                    \bigl[\nabla_1 J_\alpha(\mathbf r_1,\mathbf r_2)\!\cdot\!
                          \nabla_1 J_\alpha(\mathbf r_1,\mathbf r_3)\bigr]
                    \bigr|\phi_q\phi_s\phi_u\bigr\rangle, \nonumber \\[4pt]
    P_{pr}^{qs} &= \Bigl\langle\phi_p\phi_q\Bigl|
                  \mathrm P_{12}^{(1)}
                  \Bigl[\Pi_{12}^1(\hat H_{\text{en}}^{\text{PP}})
                       +\tfrac12\Gamma_{1212}^1(\hat H_{\text{en}}^{\text{PP}})\Bigr]
                  \Bigr|\phi_r\phi_s\Bigr\rangle.
    \end{align}
    
    \noindent
    
with the PP commutators
\begin{align}
\Pi_{ij}^i(\hat H_{\text{en}}^{\text{PP}}) &=
  \!\Bigl[\hat H_{\text{en}}^{\text{PP}}(\mathbf r_i)J_\alpha
                  -J_\alpha\hat H_{\text{en}}^{\text{PP}}(\mathbf r_i)\Bigr],\\
\Gamma_{ijij}^i(\hat H_{\text{en}}^{\text{PP}}) &=
  \hat H_{\text{en}}^{\text{PP}}(\mathbf r_i)J_\alpha^2
 -2J_\alpha\,\hat H_{\text{en}}^{\text{PP}}(\mathbf r_i)J_\alpha
 +J_\alpha^2\,\hat H_{\text{en}}^{\text{PP}}(\mathbf r_i),
\end{align}
and symmetric pemutation operators $\mathrm{P}^1_2$ and $\mathrm{P}^1_{2,3}$:
\begin{align}
    \begin{aligned}
        \mathrm{P}^1_2f(1,2)&=f(1,2)+f(2,1) \\
        \mathrm{P}^1_{2,3}f(1,2,3)&=f(1,2,3)+f(2,1,3)+f(3,2,1)
    \end{aligned}
\end{align} 
and \(\hat H_{\text{en}}^{\text{PP}}\) denotes the electron–nucleus PP energy operator. Above, we have grouped the Jastrow factor parameters $\alpha_u,\alpha_\chi,\alpha_f$ into just $\alpha$ for clarity and left the electron and nucleus position dependencies out of the equations. 

\paragraph{Density matrices.}
The antisymmetrized two-body RDM is reconstructed from the 1-RDM as  
\begin{equation}
  \gamma^{pq}_{rs}= \gamma^p_q\gamma^r_s-\gamma^p_s\gamma^r_q.
\end{equation}

(Incorporation of explicit 2-RDMs from FCIQMC calculations is under active investigation.)

\subsection{Transcorrelated post-Hartree-Fock methods}

With $\hat{H}_\text{TC}$, we use for energy estimation coupled cluster methods xTC-CCSD(T) and xTC-CCSDT (ionization) as well as xTC-FCIQMC (spin gaps \& excitations). With these methods we find the solution for the right eigenvector of the Schrödinger equation $\hat{H}_{\text{TC}} \Psi = E \Psi$ with the energy $E$ and the wave function $\Psi$. For xTC-CCSD(T) we do the calculations with a pseudocanonical $\Lambda$CCSD(T) approach using bi-orthogonal orbitals~\cite{schraivogel2021,schraivogel2023, kats2024}. 

With xTC-FCIQMC, we can directly target the ground state using imaginary time projection. This applies also to excited states that are lowest-energy states with given $S_z$. In these cases the energy can be calculated with a projected energy estimator,
\begin{equation}
  \label{eq:projected_energy_estimator}
  E(\tau)= \sum_{\mathbf{j} } \langle D_\mathbf{j} | \hat{H}_{\text{TC}} | D_\mathbf{0}\rangle \frac{c_\mathbf{j}(\tau)}{c_\mathbf{0}(\tau)},
\end{equation}
where $D_\mathbf{0}$ is the reference determinant (or largest-weight determinant in xTC-FCIQMC). However, the $4s^2 3d^{m-2} \to 4s^1 3d^{m-1}$ excitations in Mn, Fe, Co, Ni, and Cu conserve $S_z$. With these atoms, to capture the excitation energy, we propagate multiple replicas of wave functions in imaginary time while simultaneously orthogonalizing them to each other. This is done by propagating a wave function of the form 
\begin{equation}
    \label{eq:fciqmc orthogonalized}
    \Psi^\text{n}(\tau + \Delta\tau) = \hat{O}^\text{n} (\tau + \Delta\tau )\left[ \mathrm{1} -\Delta \tau\left(\hat{H}-S\mathrm{1}\right)\right]\Psi^\text{n}(\tau)
\end{equation}
with the projection operator 
 \begin{equation}
    \hat{O}^\text{n} = \mathrm{1}-\sum_{\text{m}<\text{n}}\frac{|\Psi^\text{m}(\tau)\rangle\langle\Psi^\text{m}(\tau)|}{\langle\Psi^\text{m}(\tau)|\Psi^\text{m}(\tau)\rangle}
 \end{equation}
 so that for each n, m lower states are orthogonalized to the n:th state~\cite{blunt2015}. This process does not provide the correct excited state for the right eigenvector obtained with non-hermitian TC Hamiltonian, since the right eigenvectors in general are not orthogonal. However, the projections do produce linearly independent states, and since the similarity transformation preserves the energy eigenvalues, we can still use the shift operator in Eq.~\ref{eq:fciqmc orthogonalized} to estimate the energy of the excited state, as demonstrated in a TC study of the Hubbard model ~\cite{dobrautz2019}. The shift is updated during the xTC-FCIQMC simulation to keep the population of walkers stable according to the formula
\begin{equation}
    S(\tau)=S(\tau-A\delta\tau)-\frac{\kappa}{A\delta\tau}\text{ln}\frac{N_\text{w}(\tau)}{N_\text{w}(\tau-A\delta\tau)}). \\
\end{equation}
In the above, $\kappa$ is a damping parameter, and $N_\text{w}(\tau)$ is the number of walkers at time $\tau$~\cite{booth2009}. The shift is updated every $A$ steps. We used $A=30$ for the calculations with orthogonalized replicas. This is larger than the default but proved to provide improved statistics for reblocking analysis of the stochastic energy data.

The (xTC)-FCIQMC calculations are done with spin projection $S_z$ fixed. xTC-FCIQMC simulates $j$-averaged states as we do not include spin-orbit coupling in the Hamiltonian. Hence the orbital angular momentum quantum number $L$ as given by term symbols in table \ref{table:target_occupations} gives the degeneracy of a configuration. The ground state $L$ then determines the minimal number of replicas $2L+2$ needed in the orthogonalized replica xTC-FCIQMC simulation to capture the first excited state. In practice we found that often even with only $2$ replicas, the first with $\Psi(\tau=0)$ corresponding to the ground-state configuration and the second corresponding to the excited state, the imaginary time propagation of the second replica converged to the excited state. In these cases, we did the cheaper $2$-replica calculation. However, for Fe we found almost every combination of $2$ replicas to collapse to the ground state, and hence we used $6$ replicas. For Co, a simulation with $1e^6$ walker population with aug-cc-pVTZ basis collapsed to the ground state, and an $8$-replica calculation converged the $8$th replica to the excited state (see Supplementary Material), but otherwise $2$ replicas were enough. For Ni and Cu $2$ replicas were enough.

For Cr, we used a single replica and used Eq.~\ref{eq:projected_energy_estimator} to estimate the energies of the ground and excited states. Even though FCIQMC simulation targeting the excitation with $S_z=4$ risks collapse to ground state, we found the metastable excited state during the FCIQMC simulation, and it retained its stability long enough for us to estimate the energy. The alternative would have been to do a replica calculation with $S_z=4$, but this proved to be expensive as it required many replicas and large walker population. The dynamics of the FCIQMC calculations with Cr are also included in the Supplementary Material, where the metastable state used to estimate the excitation energy is shown with vertical dashed lines.

\section{Computational details}

The computational workflow is executed using PySCF~\cite{sun2020} for orbital generation, CASINO~\cite{needs2020} for VMC/Jastrow optimization, PyTCHInt~\cite{haupt2025} for transcorrelated integral construction, Elemco~\cite{elemcojl} for coupled-cluster calculations, and NECI~\cite{guther2020} for (xTC)-FCIQMC simulations.

All of the calculations – except for separately mentioned AE tests – are obtained with pseudopotentials. For atoms Sc, Ti, V, Cr, Mn, and Fe, we use energy consistent correlated electron pseudopotentials (eCEPPs)~\cite{trail2017} and for Co, Ni, Cu, and Zn we use correlation-consistent effective core potentials (ccECPs) \cite{bennett2017}, both of which have been shown to work with transcorrelated theory \cite{simula2025}. We also use basis set functions $\theta^\alpha$ of type aug-cc-pVXZ (X=D,T,Q) optimized for these specific types of pseudopotentials. 


The Jastrow factors are optimized separately for each system, basis set, and PP combination. Only in the case of direct excitation calculations with the orthogonalized replica xTC-FCIQMC simulations we used the same Jastrow factor for the ground and excited states. The cutoffs used for the $u$, $\chi$, and $f$ terms were $4.5$, $1$, and $1$, respectively. We provide the optimized Jastrow factors in aug-cc-pVTZ and aug-cc-pVQZ basis sets in the Supplementary Material. Also in the Supplement, we include the time steps $\Delta\tau$ used in the xTC-FCIQMC simulations, which are optimized in the equilibration phase. 

\section{Results}

\subsection{Orbitals and wave function compactification}

\begin{figure}
      \centering
    \includegraphics[scale=.62]{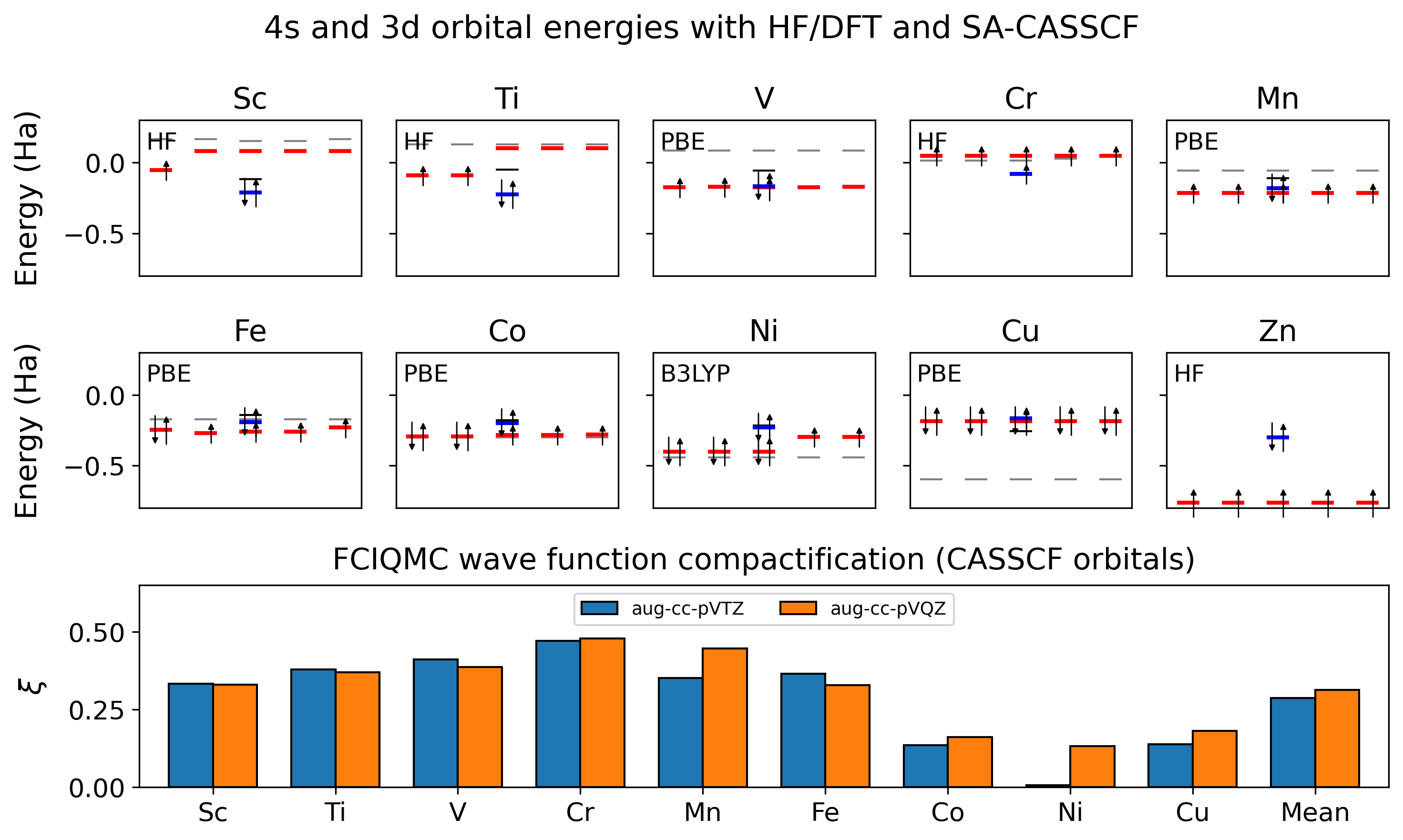}
    \caption{Energies of orbitals 4s and 3d, obtained with SA-CASSCF (4s black and 3d grey) and HF/DFT methods (4s is blue and 3d red). The orbital occupation is denoted with arrows. Below the orbital energies are histograms of the wave function compactification coefficients $\xi$ from FCIQMC and xTC-FCIQMC simulations, using  SA-CASSCF orbitals.}
    \label{fig:multireference}
\end{figure}

Figure \ref{fig:multireference} shows the energies of the 4s and 3d orbitals for the transition metals Sc--Zn, obtained with the SA-CASSCF and HF/DFT methods. The orbital occupation from HF or DFT simulation is also shown. With SA-CASSCF, we always find degenerate 3d orbitals, with fractional occupation. Without SA-CASSCF, we used DFT in cases where HF could not produce a degenerate 3d shell nor find the correct reference as denoted from the literature. Sc and Ti were exceptions, as we could not find degenerate 3d orbitals with neither HF nor DFT with any functional tried. We chose to use HF for Sc and Ti. We also used HF with Cr and Zn, as these atoms are half-filled and fully-filled, respectively, and hence the HF method provides a good reference. For the rest of the atoms we found DFT to provide a better reference, with degenerate d shell and correct occupation. An exception was Ni, where finding a solution proved difficult with HF and DFT. In general we used the PBE functional with DFT, but for Ni we used B3LYP, which was slightly better in this case.

Figure~\ref{fig:multireference} also shows the histograms of the wave function compactification coefficients $\xi$ from FCIQMC and xTC-FCIQMC simulations, using the SA-CASSCF orbitals, in AVTZ and AVQZ basis sets. The compactification can be quantified \cite{haupt2023} as 
\begin{equation}
    \xi = \frac{c^{\text{TC}}_\mathbf{0}-c_\mathbf{0}^\text{non-TC}}{1-c_\mathbf{0}^\text{non-TC}},
\end{equation}
where $c_\mathbf{i}$ are the FCIQMC determinant coefficients, and $c_\mathbf{0}$ is the coefficient of the largest-weight determinant. The compactification coefficients $\xi$ are a measure of the compactification of the wave function due to TC, as compared to non-TC. $\xi$ refers to the proportion of correlations included by the similarity transformation of the Hamiltonian into the reference determinant. A value of $\xi=0$ means that the wave function is not compactified at all, and a value of $\xi=1$ means that the wave function is fully described by the reference determinant.  Note that $\xi$ can also theoretically be negative, indicating a less compact wavefunction compared to the non-TC wavefunction, but this is not observed for the systems we studied, indicating that the TC method in practice always leads to compactification of the right-eigenvector, albeit by varying amounts, depending on the system. 

Based on $\xi$ in Fig.~\ref{fig:multireference}, we can see that the compactification is large, of the order of $0.35$-$0.49$ for atoms Sc--Fe.  For Co, Ni, and Cu, the compactification is significantly smaller, ranging between $0.01$ and $0.19$. The differences between the aug-cc-pVTZ and aug-cc-pVQZ basis sets are small. On average, $\xi$ is $\sim0.3$, signifying a strong concentration of the correlation effects into the reference determinant. These values also mean that lower levels of excitations are usually required to describe the wave function. Hence FCIQMC gets cheaper and coupled cluster methods get more accurate with the TC Hamiltonian.

\subsection{Total energies}

\begin{figure}
  \centering
  \includegraphics[width=\textwidth]{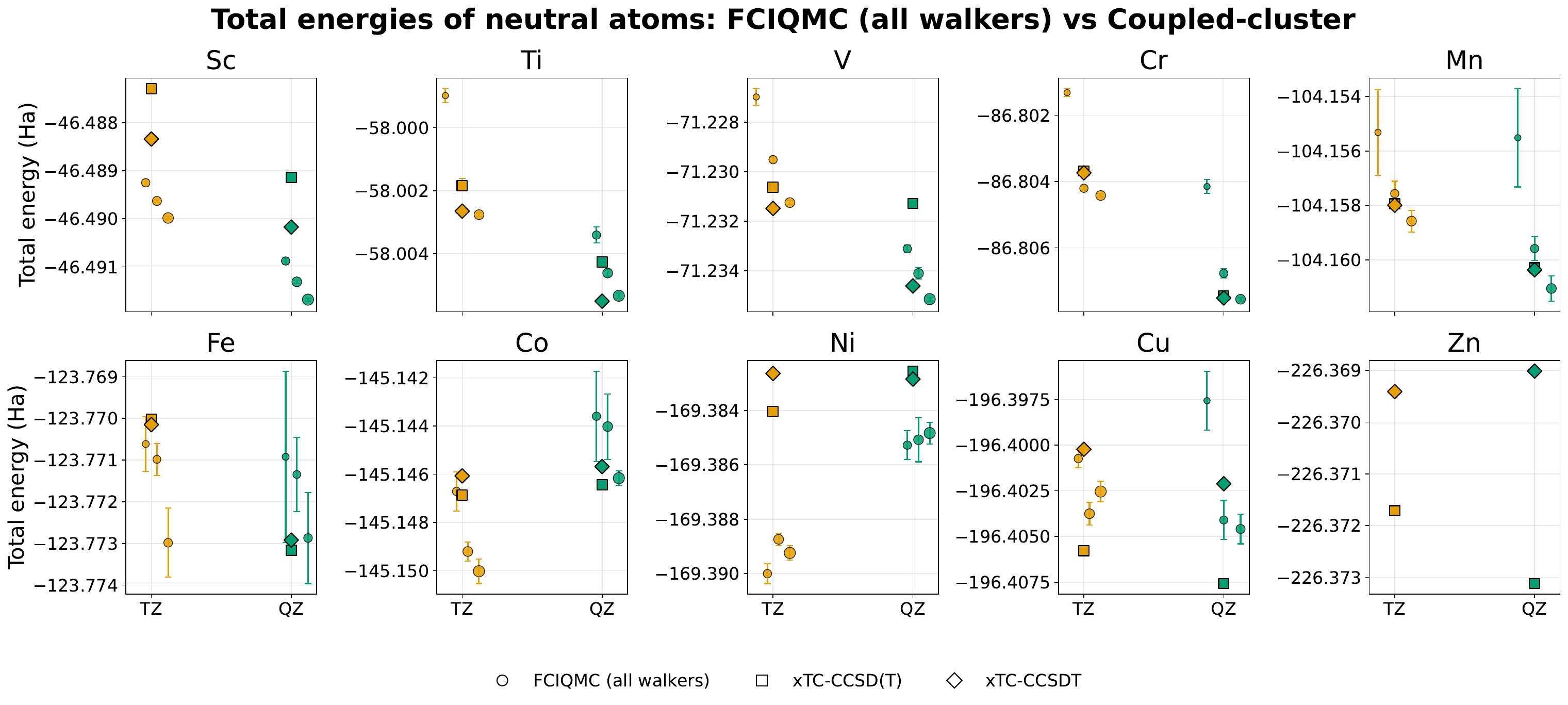}
  \caption{Total energies of the transition metals Sc--Zn, calculated with xTC-CCSD(T) (squares), xTC-CCSDT (diamonds), xTC-FCIQMC (circles) in AVTZ (orange) and AVQZ (green) basis sets. The energies are given in Hartree. The error bars show the statistical uncertainty of the energy estimates (in some cases the marker is larger than the errorbar). Multiple xTC-FCIQMC energies are shown for a given element and basis set, corresponding to different walker populations. The larger the walker population, the larger the marker, with walker populations used shown in Table~\ref{tab:walkers}. }
  \label{fig:total_energies}
\end{figure}

Figure~\ref{fig:total_energies} shows the total energies of the transition metals Sc--Zn, calculated with xTC-CCSD(T), xTC-CCSDT, and xTC-FCIQMC with varzing walker numbers in AVTZ and AVQZ basis sets. The energies are given in Hartree, and the error bars show the statistical uncertainty of the FCIQMC energy estimates. The AVQZ basis xTC-CCSDT energies are found to be closer than $2$mHa to the xTC-FCIQMC energies. Given that the  xTC-CCSDT method is a single-reference method, with orbitals taken from HF or DFT, this is a good result, as it shows that these often highly multireference systems are accessible with coupled cluster when transcorrelated Hamiltonian is used. Also, given the close correspondence in total energies of these two very different TC approaches (HF or DFT orbitals and CC vs. SA-CASSCF orbitals and FCIQMC), we can conclude that the TC method is robust to the choice of orbitals, at least in these cases.

\subsection{4s to 3d Excitations}

\begin{figure}[H]
    \centering
    \includegraphics[scale=.32]{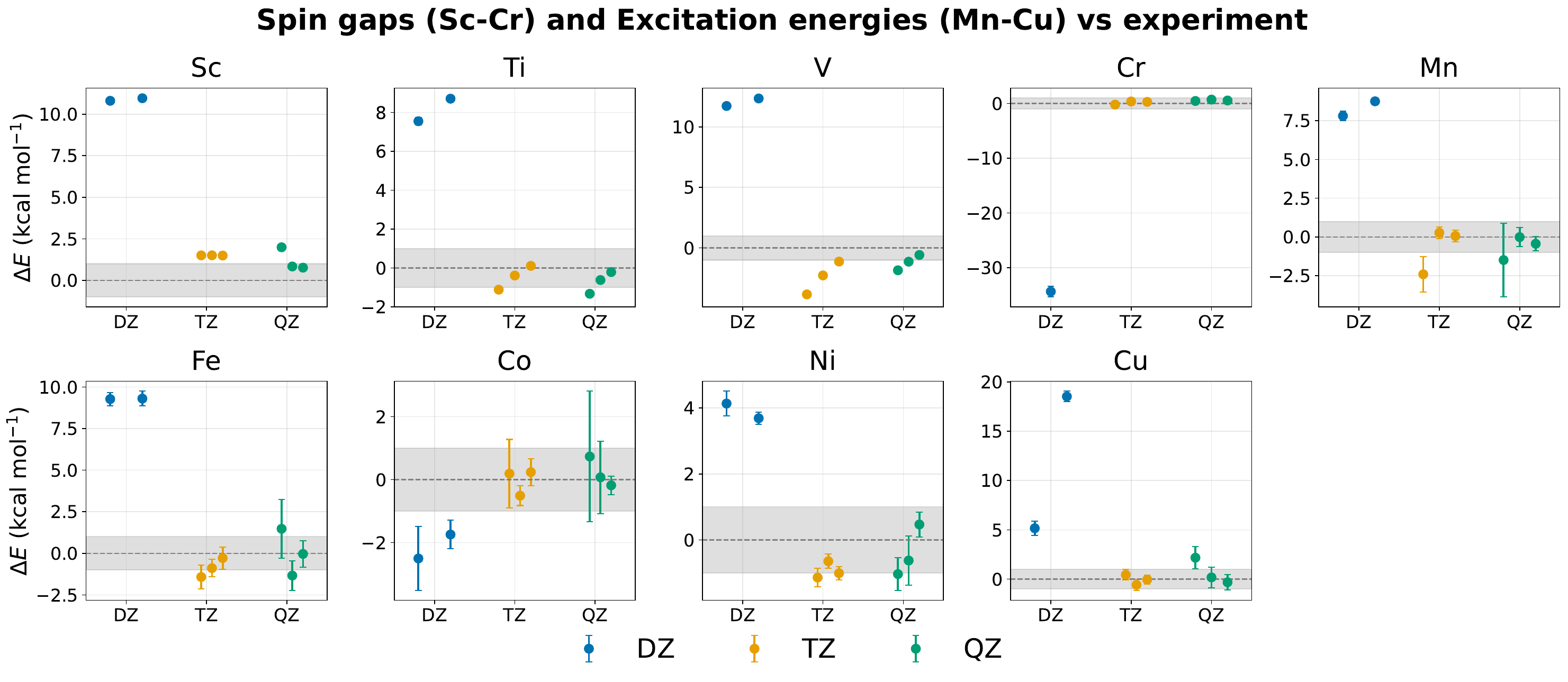}
    \caption{Errors in $4s \rightarrow 3d$ electronic excitation energies of atoms Sc--Cu, evaluated with xTC-FCIQMC against experimental values, $\Delta E = E_{\mathrm{Theory}} - E_{\mathrm{Expt.}}$. Multiple basis sets (AVDZ, AVTZ, AVQZ) are shown. 
    For each element and basis set, we plot a number of results with varying walker populations (increasing from left to right), with the exact walker population sizes shown in Table~\ref{tab:walkers}.
    The grey region denotes chemical accuracy 
    ($\pm 1\,\mathrm{kcal/mol}$).}
    \label{fig:excitation_energies}
\end{figure}

Figure~\ref{fig:excitation_energies} illustrates the differences between the xTC-FCIQMC $4s \to 3d$ transition energies and ($j$-averaged) experimental values in transition metals from Sc to Cu. Spin-gap excitations (Sc–Cr) and spin-conserving excitations (Mn–Cu) were both examined. The gaps with different walker populations are shown for each basis set. Table \ref{tab:walkers} shows the walker populations used in the xTC-FCIQMC calculations for each element and basis set.

\begin{table*}[tbp]
  \centering
  \caption{Walker populations ($N_{\mathrm w}$) employed in the imaginary-time
           xTC–FCIQMC simulations.  All calculations use
           aug-cc-pV\emph{X}Z basis sets (\emph{X}=D, T, Q).}
  \label{tab:walkers}

  \small \renewcommand\arraystretch{1.05}

  \begin{tabularx}{\textwidth}{@{}l
      >{\raggedright\arraybackslash}p{3.4cm}
      >{\raggedright\arraybackslash}p{3.8cm}
      >{\raggedright\arraybackslash}p{3.8cm}@{}}
    \toprule
    Atom &
    aug-cc-pVDZ &
    aug-cc-pVTZ &
    aug-cc-pVQZ \\
    \midrule
    Sc & \numlist{2e6;1e7}
        & \numlist{5e6;1e7;5e7}
        & \numlist{1e6;5e6;2e7} \\
    Ti & \numlist{1e6;5e6}
        & \numlist{1e6;5e6;2e7}
        & \numlist{5e6;2e7;1e8} \\
    V  & \numlist{1e6;5e6}
        & \numlist{1e6;5e6;2e7}
        & \numlist{5e6;2e7;1e8} \\
    Cr & \num{1e6}
        & \numlist{1e6;5e6;2e7}
        & \numlist{1e6;5e6;2e7} \\
    Mn & \numlist{1e6;5e6}
        & \numlist{1e6;5e6;2e7}
        & \numlist{1e6;5e6;2e7} \\
    Fe & \numlist{1e6;5e6}
        & \numlist{5e6;1e7;2e7}
        & \numlist{5e6;1e7;2e7} \\
    Co & \numlist{1e6;5e6}
        & \numlist{5e6;2e7;1e8}
        & \numlist{5e6;2e7;1e8} \\
    Ni & \numlist{1e6;5e6}
        & \numlist{5e6;2e7;1e8}
        & \numlist{5e6;2e7;1e8} \\
    Cu & \numlist{1e6;5e6}
        & \numlist{5e6;2e7;1e8}
        & \numlist{5e6;2e7;5e7} \\
    \bottomrule
  \end{tabularx}
\end{table*}

The results show without exception that a double-zeta basis is insufficient for accurate transcorrelated calculations. However, in all of the cases we are in the chemically accurate regime already with the triple-zeta basis set, with only minor changes introduced by the quadruple-zeta basis set. 

As can be seen from Table ~\ref{tab:walkers} and Fig.~\ref{fig:excitation_energies}, we need more walkers for V, Co, Ni, and Cu than for the rest of the systems. This is to be expected, as these systems are more strongly correlated and the significant determinants are more spread out in the Hillbert space, requiring more walkers to give accurate weights for all of the determinants. Ti is not generally considered particularly challenging, and we found the use of $1e8$ walkers to be excessive, as results with $2e7$ walkers were already chemically accurate. We also want to point out that the use of more walkers in some cases would probably improve the results. However, the computational cost of pushing FCIQMC simulations with more than $1e8$ walkers to low levels of stochastic noise is not worth the effort; the results are already chemically accurate and it is very unlikely that the results would change significantly with more walkers. 


\subsection{Ionization Potentials}
\begin{figure}[H]
    \centering
    \includegraphics[scale=.32]{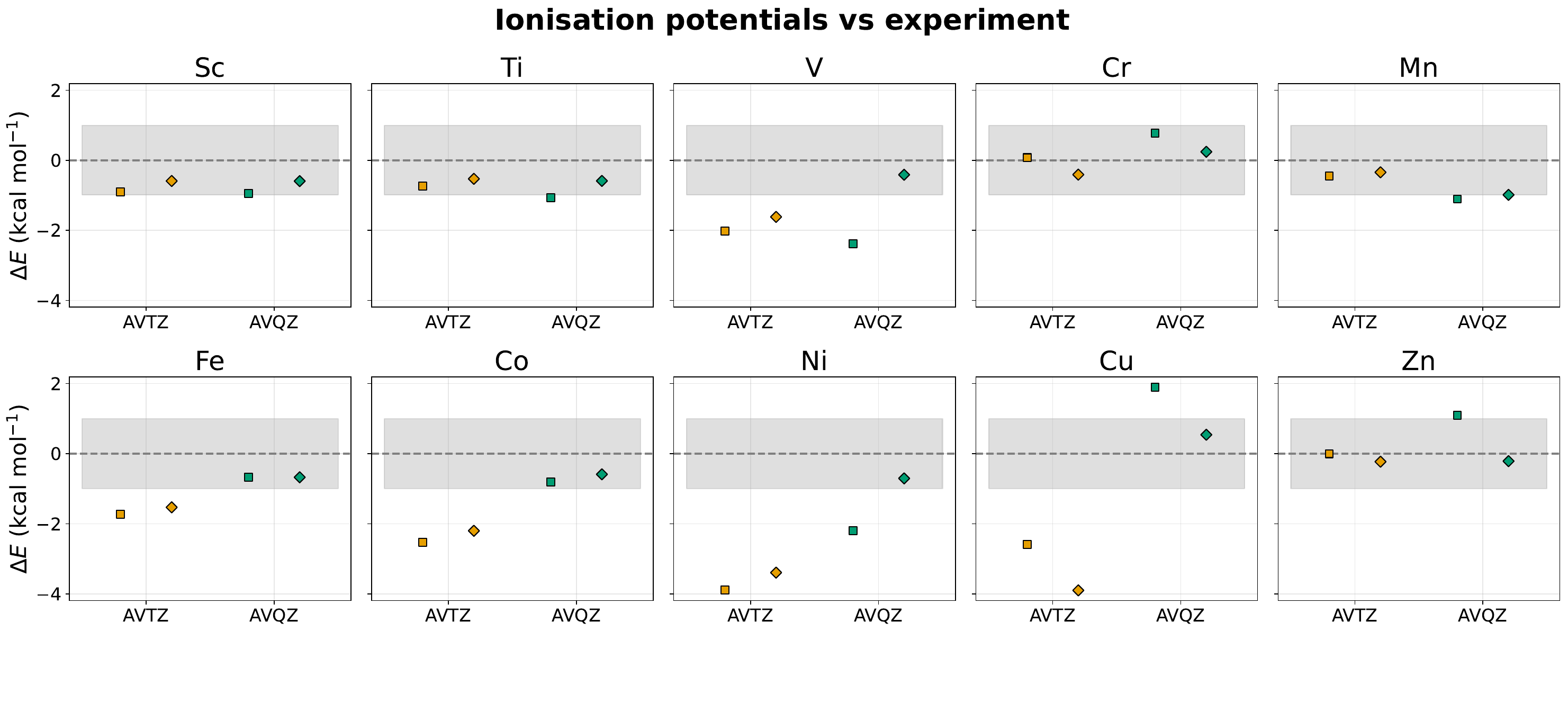}
    \caption{Errors in theoretical ionization potentials against experimental 
    values for the transition metals, $\Delta E = E_{\mathrm{Theory}} - E_{\mathrm{Expt.}}$. 
    xTC-CCSD(T) (rectangle) and xTC-CCSDT (diamond)
     results are shown for aug-cc-pVTZ (yellow) and aug-cc-pVQZ  (green)
    basis sets. Grey region denotes chemical accuracy.}
    \label{fig:ionization_potentials}
\end{figure}

The discrepancy of ionization potentials computed with xTC-CCSD(T) and xTC-CCSDT against experiment are shown in Figure~\ref{fig:ionization_potentials}. We present again results with both methods in AVZT and AVQZ basis sets, without studying the double-zeta basis set, which proved to be inaccurate in the previous section. Here we also find chemical accuracy with every element, when AVQZ basis and xTC-CCSDT is used. CCSDT offers a small, but in some cases important improvement over CCSD(T) for the ionization potentials. Especially for V, Co, Ni, and Cu the difference is significant, as CCSDT is chemically accurate with AVQZ while CCSD(T) is not.

\subsection{TC accuracy for transition metal atoms}

Table \ref{tab:ex_ip_allbasis} summarizes the excitation energies for the $4s^2 3d^{m-2} \to 4s^1 3d^{m-1}$ transitions, as well as the first ionization potentials, calculated with xTC-FCIQMC and xTC-CCSD(T)/xTC-CCSDT. With xTC-FCIQMC we used the largest walker number result for each atom and basis set. The results are compared to experimental values, with deviations noted. The mean squared error (MSE) and mean absolute error (MAE) of the excitation energies and ionization potentials are also shown in Table \ref{tab:ex_ip_allbasis} for each basis set and method. The experimental values used were averages over different $j$-states. 
\begin{table*}[htbp]
    \centering\vspace{-2cm}
    \small
    \setlength{\tabcolsep}{6pt}
    \renewcommand{\arraystretch}{1.15}
    \caption{4\(s\!\rightarrow\!3d\) excitation energies (xTC‐FCIQMC) and first‑ionisation
             potentials [xTC‐CCSD(T) (pT) and xTC‐CCSDT (T)] of the first‑row transition metals.
             All energies are in kcal\,mol$^{-1}$; statistical uncertainties for the
             xTC-FCIQMC data are 1 s.e. in parentheses.
             Deviations are $\Delta = E_{\text{theory}}-E_{\text{exp}}$.
             Experimental references are from Ref.~\cite{moore1971} as presented in
            Refs.~\cite{blabanov2005,blabanov2006}. for $4s^2 3d^{m-2} \to 4s^1 3d^{m-1}$ excitations and from Refs.~\cite{ip1_sugar1985,ip2_ralph1990,ip3_sohl1990,ip4_james1990,ip5_kramida2013} for ionization potentials.  For Sc and Zn, the AE ionization potential results are shown, evaluated in aug-cc-pVTZ basis (AE TZ), where post-HF calculations were done with the neon core frozen. In the parentheses we show the 
            AE results with the relativistic corrections, estimated at CBS limit with high-accuracy coupled cluster and FCI corrections from \cite{blabanov2006}.}
    \label{tab:ex_ip_allbasis}
    \begin{tabular}{@{}l l
                    r r r
                    r r r r r@{}}
      \toprule
      \multirow{2}{*}{Element} & \multirow{2}{*}{Basis} &
        \multicolumn{3}{c}{Excitation \(4s\rightarrow3d\)} &
        \multicolumn{5}{c}{Ionisation potential} \\[-0.3em]
        & & Th. & Exp. & \(\Delta\) &
            Th.\,(pT) & Th.\,(T) & Exp. & \(\Delta\)(pT) & \(\Delta\)(T) \\
      \cmidrule(lr){3-5}\cmidrule(lr){6-10}
      \multirow{3}{*}{Sc}
        & DZ & 43.87(2) & \multirow{3}{*}{32.91} & +10.96
             & ---      & ---      & \multirow{3}{*}{151.306} & ---   & ---   \\
        & TZ & 34.36(5) &          &  +1.45
             & 150.402  & 150.715  &                        & -0.90 & -0.59 \\
        & QZ & 33.40(1) &          &  +0.49
             & 150.162  & 150.501  &                        & -1.14 & -0.81 \\
         & AE TZ & --  &  --        &  
              & 153.786  & 149.878  &                        & 2.48 (3.23) & -1.43 (-0.68) \\
      \midrule
      \multirow{3}{*}{Ti}
        & DZ & 27.29(4) & \multirow{3}{*}{18.58} &  +8.71
             & ---      & ---      & \multirow{3}{*}{157.455} & ---   & ---   \\
        & TZ & 18.68(4) &          &  +0.10
             & 156.723  & 156.925  &                        & -0.73 & -0.53 \\
        & QZ & 18.47(5) &          &  -0.11
             & 156.540  & 156.897  &                        & -0.92 & -0.56 \\
      \midrule
      \multirow{3}{*}{V}
        & DZ & 18.00(4) & \multirow{3}{*}{ 5.65} & +12.35
             & ---      & ---      & \multirow{3}{*}{155.235} & ---   & ---   \\
        & TZ &  4.56(5) &          &  -1.09
             & 153.218  & 153.620  &                        & -2.02 & -1.62 \\
        & QZ &  5.10(1) &          &  -0.55
             & 153.384  & 153.820  &                        & -1.85 & -1.42 \\
      \midrule
      \multirow{3}{*}{Cr}
        & DZ & -11(1)   & \multirow{3}{*}{23.13} & -34.34
             & ---      & ---      & \multirow{3}{*}{156.025} & ---   & ---   \\
        & TZ & 23.40(4) &          &  +0.27
             & 156.099  & 155.616  &                        & +0.07 & -0.41 \\
        & QZ & 23.67(7) &          &  +0.54
             & 156.830  & 156.299  &                        & +0.81 & +0.27 \\
      \midrule
      \multirow{3}{*}{Mn}
        & DZ & 58.2(2)  & \multirow{3}{*}{49.47} &  +8.73
             & ---      & ---      & \multirow{3}{*}{171.464} & ---   & ---   \\
        & TZ & 49.6(4)  &          &  +0.13
             & 171.018  & 171.121  &                        & -0.45 & -0.34 \\
        & QZ & 49.5(6)  &          &  +0.03
             & 170.554  & 170.590  &                        & -0.91 & -0.87 \\
      \midrule
      \multirow{3}{*}{Fe}
        & DZ & 29.5(4)  & \multirow{3}{*}{20.18} &  +9.32
             & ---      & ---      & \multirow{3}{*}{182.253} & ---   & ---   \\
        & TZ & 19.9(7)  &          &  -0.28
             & 180.521  & 180.722  &                        & -1.73 & -1.53 \\
        & QZ & 20.0(6)  &          &  -0.18
             & 181.932  & 182.040  &                        & -0.32 & -0.21 \\
      \midrule
      \multirow{3}{*}{Co}
        & DZ &  7.9(5)  & \multirow{3}{*}{ 9.62} &  -1.72
             & ---      & ---      & \multirow{3}{*}{181.453} & ---   & ---   \\
        & TZ &  9.6(5)  &          &  -0.02
             & 178.919  & 179.253  &                        & -2.53 & -2.20 \\
        & QZ &  9.0(4)  &          &  -0.62
             & 179.691  & 179.911  &                        & -1.76 & -1.54 \\
      \midrule
      \multirow{3}{*}{Ni}
        & DZ &  3.0(2)  & \multirow{3}{*}{-0.69} &  +3.69
             & ---      & ---      & \multirow{3}{*}{175.103} & ---   & ---   \\
        & TZ & -1.7(2)  &          &  -1.01
             & 171.211  & 171.710  &                        & -3.89 & -3.39 \\
        & QZ & -0.9(5)  &          &  -0.21
             & 172.908  & 174.399  &                        & -2.196 & -0.705 \\
      \midrule
      \multirow{3}{*}{Cu}
        & DZ & -15.8(5) & \multirow{3}{*}{-34.37} & +18.57
             & ---      & ---      & \multirow{3}{*}{178.153} & ---   & ---   \\
        & TZ & -34.4(5) &          &  -0.03
             & 175.561  & 174.253  &                        & -2.59 & -3.90 \\
        & QZ & -34.2(6) &          &  +0.17
             & 179.856  & 178.392  &                        & +1.70 & +0.24 \\
      \midrule
      \multirow{2}{*}{Zn}
        & TZ & ---      & \multirow{2}{*}{---} & ---
             & 216.607  & 216.381  & \multirow{2}{*}{216.609} & -0.00 & -0.23 \\
        & QZ & ---      &                         & ---
             & 216.683  & 216.576  &                         & +0.07 & -0.03 \\
         & AE TZ & ---      &      ---                   & ---
              & 211.593  & 211.319  &                         & -5.02 (-0.48) & -5.30 (-0.76) \\

      \bottomrule
    \end{tabular}
  \end{table*}

  \begin{table}
  \centering
  \caption{Mean squared error (MSE) and mean absolute error (MAE) of the excitation energies and ionization potentials for the transition metals Sc--Zn, calculated with xTC-FCIQMC, xTC-CCSD(T), and xTC-CCSDT. The errors are shown for three different basis sets: aug-cc-pVDZ, aug-cc-pVTZ, and aug-cc-pVQZ. The MSE and MAE are given in kcal/mol.}
  \label{tab:ex_ip_allbasis_errors}
  \begin{tabular}{@{}l l
    r r r
    r r r r r@{}}
\toprule
    \multicolumn{10}{c}{\textbf{Error statistics (kcal\,mol$^{-1}$)}}\\
    \cmidrule(lr){1-10}
    \multicolumn{2}{l}{Excitation MSE} &
      \multicolumn{8}{l}{DZ = 683.66,\; TZ = 1.58,\; QZ = 0.60, \; Ref.~\cite{blabanov2006}: 0.21} \\[-0.2em]
    \multicolumn{2}{l}{Excitation MAE} &
      \multicolumn{8}{l}{DZ =  12.04,\; TZ =  0.52,\; QZ =  0.39,\; Ref.~\cite{blabanov2006}: 0.43} \\[0.3em]
    \multicolumn{2}{l}{Ionisation CCSD(T) MSE} &
      \multicolumn{8}{l}{DZ = ---  ,\; TZ = 3.69,\; QZ = 2.03} \\[-0.2em]
    \multicolumn{2}{l}{Ionisation CCSD(T) MAE} &
      \multicolumn{8}{l}{DZ = ---  ,\; TZ =  1.49,\; QZ =  1.30} \\[0.3em]
    \multicolumn{2}{l}{Ionisation CCSDT  MSE} &
      \multicolumn{8}{l}{DZ = ---  ,\; TZ = 3.75,\; QZ = 0.35,\; Ref.~\cite{thomas2015}: 0.18} \\[-0.2em]
    \multicolumn{2}{l}{Ionisation CCSDT  MAE} &
      \multicolumn{8}{l}{DZ = ---  ,\; TZ =  1.47,\; QZ =  0.55,\; Ref.~\cite{thomas2015}: 0.31} \\
    \bottomrule
  \end{tabular}
  \end{table}

The table shows that in a quadruple-zeta basis set the xTC-FCIQMC and the xTC-CCSDT methods deliver chemical accuracy  for all of the transition-metal atoms studied. The AVQZ xTC-FCIQMC MAE for the excitations is the best to date, although given the stochastic uncertainties, no conclusion can be drawn on whether the current calculations are more or less accurate than the results by  Blabanov and Peterson~\cite{blabanov2006}.  In any case, to obtain this level of accuracy without the need for calculating relativistic corrections or extrapolation to CBS, with a modest basis set size, is a remarkable result and a proof that the TC theory offers a way to tackle correlations in heavy atoms with less computational cost than traditional methods.

Also, Table~\ref{tab:ex_ip_allbasis} presents the most accurate ionisation potential calculation for the atoms Sc--Zn with purely coupled-cluster-based methods (without FCI corrections, for example, as in \cite{blabanov2006}), to the best of our knowledge. We do not reach the level of accuracy of the best known results, which are obtained with initiator FCIQMC~\cite{thomas2015}, but we are easily within chemical accuracy, without the need for extrapolation to CBS or relativistic corrections to the Hamiltonian and basis sets.

We performed AE ionisation potential calculations with the same methods as in this study for Sc and Zn, with the $1s$, $2s$, and $3p$ frozen in the correlated CC simulation, to assess the effect of pseudopotentials. In table \ref{tab:ex_ip_allbasis}, we show the AE results in the AVTZ basis set, where the neon core was frozen in the post-HF calculations. The AE results are chemically accurate, but not as accurate as the PP results, even with the relativistic corrections included from \cite{blabanov2006}. Because we held the neon core frozen, the orbital space used in post-HF was equal to the PP space. While we do not have a conclusive argument of why PP TC is more accurate, we list three reasons that can be responsible for this. First, with pseudopotentials we do not have the electron-nuclear cusp problem, which in the AE calculation is resolved by including the cusp condition in the Jastrow factor. This can lead to a less accurate treatment of the cusp than the PP approximation. Second, with pseudopotentials the VMC optimization of the Jastrow factor is much cheaper, but allows also a more targeted optimization of the Jastrow factor for the valence electrons. Third, the variance of the reference wave function during the VMC optimization of the Jastrow is extremely large with TM atoms. Hence it can be that the resolution of $0.1$mHa in TC energies obtained with the optimization procedure described in Ref.~\cite{haupt2023} does not hold with TM atoms.  

\section{Conclusions}

We have presented a transcorrelated study to address a traditional problem of atomic physics, namely the calculation of ionisation and excitation energies of transition metal atoms Sc--Zn. We showed the theory behind our approach, presented the workflow and methods used in the calculations, and displayed the results obtained from xTC-FCIQMC, xTC-CCSD(T), and xTC-CCSDT. We inspected differences between the use of HF, DFT and SA-CASSCF orbitals, wave function compactification, and basis set covergence with basis sets AVDZ, AVZT, and AVQZ. 

We used the PP approximation throughout the study. With already success in lighter atoms~\cite{simula2025}, this study augments the applicability of the TC method with pseudopotentials to heavy atoms, and hence opens up interesting possibilities for tackling more challenging systems. 

We reached chemical accuracy against experiment in terms of both MAE and MSE for spin gaps and excitations when using an AVQZ basis and xTC-FCIQMC. The MAE obtained in this study is the current state-of-the-art, very close to previous state-of-the-art coupled cluster calculations \cite{blabanov2006}. Most of the results were converged in the AVZT basis already, with MAE and MSE of $0.52$ and $1.58$ kcal/mol$^{-1}$, respectively, which means that we do not need extrapolation to the CBS limit with TC theory.

For ionisation energies, we obtained chemical accuracy for atoms Sc--Zn using xTC-CCSDT and an AVQZ basis set. The MAE and MSE of our best TC results in the AVQZ basis were $0.55$ and $0.35$ kcal/mol$^{-1}$, respectively. The results were obtained using only coupled cluster-based methods, without FCI corrections or DK theory. 

With the current results at hand, the TC theory can be applied to more complex systems containing transition metals, such as transition-metal complexes or strongly correlated solid-state systems. When moving to these challenging systems, active space calculations or embedding models could be employed to circumvent the limitations due to drastic increase in computational cost with increasing number of electrons. In any case, the TC theory provides a promising route to tackle strongly correlated systems with less computational cost than traditional methods.


\end{document}